\documentclass[aps,prl,notitlepage,nofootinbib,showpacs,twocolumn]{revtex4-1}

\usepackage{graphicx}
\usepackage{amssymb}
\usepackage{amsmath}
\usepackage{bm}
\usepackage{latexsym}
\usepackage{epsfig}
\usepackage{psfrag}
\usepackage{ulem}
\usepackage{color}

\def\nn{\nonumber} 
\def\pa{{\partial}}
\def\f{\frac}
\def\l{\left}
\def\r{\right}
\def\d{{\rm d}}
\def\Mp{M_{_{\rm Pl}}}

\newcommand{\viz}{\textit{viz.~}}

\begin{document}

\title{Primordial features from ekpyrotic bounces}
\author{Rathul Nath Raveendran$^\dag$ and L.~Sriramkumar$^\ddag$}
\affiliation{$^\dag$The Institute of Mathematical Sciences, HBNI, CIT Campus, 
Chennai~600113, India\\
$^\ddag$Department of Physics, Indian Institute of Technology Madras, 
Chennai~600036, India}
\begin{abstract}
Certain features in the primordial scalar power spectrum are known 
to improve the fit to the cosmological data.
We examine whether bouncing scenarios can remain viable if future 
data confirm the presence of such features.
In inflation, the fact that the trajectory is an attractor permits 
the generation of features.
However, bouncing scenarios often require fine tuned initial conditions, 
and it is {\it only}\/ the ekpyrotic models that allow attractors.
We demonstrate, {\it for the first time},\/ that ekpyrotic scenarios 
can generate specific features that have been considered in the context 
of inflation.
\end{abstract}
\maketitle


\noindent
\underline{\it Inflation, features and bounces:}~The precise 
observations of the anisotropies in the Cosmic Microwave 
Background (CMB) by WMAP and Planck~\cite{wmap,planck} point 
to a primordial scalar power spectrum that is nearly independent 
of scale and is largely adiabatic~\cite{ccp}.
The most popular paradigm to generate perturbations of such nature 
is the inflationary scenario~\cite{ci}.
As is well known, inflation is driven by scalar fields (see, for instance,
the reviews~\cite{ir}).
There exist many models which permit inflation of the slow roll type 
leading to power spectra that are consistent with the cosmological data 
(for a comprehensive list, see Ref.~\cite{cl}).

\par

Though a nearly scale invariant primordial power spectrum as
generated by slow roll inflation is consistent with the observational
data, there have been repeated (model dependent as well as model 
independent) efforts to examine if the power spectrum contains 
features~\cite{ci,fr}.
It has been found that certain features improve the fit to the 
CMB data~\cite{ci,fr,lpq,flm,oip,atf,rcps}.
The possibility of such features gains importance because of the 
reason that, if they are confirmed by future observations, they 
can strongly limit the space of viable models.
While such features can be produced in inflationary models which 
permit deviations from slow roll~\cite{lpq,flm,oip,atf}, it is
imperative to examine if they can also be generated in alternative
scenarios.

\par

Even as inflation has been remarkably successful, alternative
scenarios have been explored for the origin of primordial 
perturbations. 
Amongst these alternatives, the most investigated are the classical 
bouncing scenarios (for recent reviews, see Refs.~\cite{br}).
Recall that, the primary goal of the inflationary paradigm is to overcome 
the horizon problem and provide natural initial conditions for the 
perturbations when they are well inside the Hubble radius during the early
stages of the accelerated expansion.
The bouncing scenarios can permit similar initial conditions to be imposed
on the perturbations during the contracting phase, provided the early phase 
is undergoing {\it decelerated}\/ contraction.
More than a handful of bouncing scenarios have been constructed that 
result in primordial power spectra that are consistent with the 
observations (see, for instance, Refs.~\cite{vb}).

\par

It is rather straightforward to construct a model of inflation and,
as we mentioned, there exist many models of inflation that perform 
well against the cosmological data.
In contrast, it proves to be an intricate task to construct bouncing 
models that are free of pathologies (for a list of difficulties faced, 
see, for example, Refs.~\cite{br,dwb}).
Moreover, the inflationary trajectory is almost always an attractor, 
which permits inflation to be achieved easily.
However, bouncing scenarios often require fine tuned initial conditions.
It is the attractor nature of the inflationary trajectory which allows 
for the generation of features in the primordial spectrum through brief
periods of deviation from slow roll.
The fact that the trajectory is an attractor ensures that slow roll 
inflation is restored after such departures. 
The fine tuned conditions required for bouncing scenarios implies that 
features cannot be generated in these models.
For instance, near matter bounces, which can be easily constructed, do 
not behave as attractors and hence they cannot return to the original
trajectory if departures are introduced~\cite{levy-2017}.
This implies that such models will be ruled out if cosmological data 
confirm the presence of features in the primordial spectrum.

\par

Amongst the bouncing models, it is {\it only}\/ 
the ekpyrotic scenario that permits trajectories which are 
attractors (for the original ideas, see Refs.~\cite{em}; for more
recent discussions, see Refs.~\cite{em-rd}).
Another advantage of the ekpyrotic model is the fact that the
anisotropic instabilities which may arise during the contracting 
phase can be suppressed since the energy density of the ekpyrotic 
source dominates the evolution.
However, ekpyrotic models driven by a single scalar field generate
spectra of curvature perturbations that have a strong blue tilt.
Therefore, models involving more than one field are considered, with
the ekpyrotic contracting phase being dominated by isocurvature 
perturbations with a nearly scale invariant spectrum. 
The second field is utilized to convert the isocurvature perturbations 
into adiabatic perturbations, eventually resulting in a nearly scale 
invariant curvature perturbation spectrum as is required by the 
observations (see, for example, Refs.~\cite{fertig-2016,ng-in-em}).
In this work, {\it for the first time},\/ we examine if features 
can be generated in the curvature perturbation spectrum in 
ekpyrotic bounces.
We shall explicitly construct ekpyrotic potentials which permit the 
generation of features that have been considered in the context of 
inflation.

\par

We shall set $\hbar=c=1$, $\Mp=1/\sqrt{8\,\pi\,G}$, and work with 
the metric signature~$(-,+,+,+)$.
Also, as usual, an overdot and an overprime shall denote derivatives 
with respect to the cosmic time~$t$ and the conformal time~$\eta$.


\vskip 6pt\noindent
\underline{\it Ekpyrotic attractor:}~We shall first briefly discuss 
the dynamics of the background in an ekpyrotic model, specifically 
showing that a negative definite potential for the scalar field 
admits an attractor during the contracting phase.
The model we shall consider involves two scalar fields $\phi$ and 
$\chi$, which are governed by the following action consisting of 
the potential $V(\phi,\chi)$ and a 
function~$b(\phi)$~\cite{em-rd,lalak-2007}:
\begin{eqnarray} 
S[\phi,\chi]
&=&\int\d^{4}x\, \sqrt{-g}\,\biggl[-\f{1}{2}\,\pa_\mu\phi\,
\pa^\mu\phi\nn\\
& &-\f{{\rm e}^{2\, b(\phi)}}{2}\,\pa_\mu \chi\,\pa^\mu\chi
-V(\phi,\chi)\biggr].
\end{eqnarray}
We shall work with the potential $V(\phi,\chi)= V_{\rm ek}(\phi)=
V_0\, {\rm e}^{\lambda\, \phi/\Mp}$ and choose $b(\phi)=\mu\, 
\phi/(2\, \Mp)$, where $\lambda$ and $\mu$ are positive constants. 
To examine the stability of the background, it is convenient 
to write the background equations in terms of the following 
dimensionless variables~\cite{em-rd}:
\begin{equation}
(x,y,z)\equiv \l(\f{\dot{\phi}}{\sqrt{6}\, H\, \Mp}, 
\f{\sqrt{V}}{\sqrt{3}\,H\, \Mp},\f{{\text e}^b\dot{\chi}}{\sqrt{6}\,H\, 
\Mp}\r). 
\end{equation}
In terms of the variables $(x,y,z)$, the equations governing the two 
scalar fields can be written as 
\begin{subequations}
\begin{eqnarray}
\f{\d x}{\d N}
&=& -3\, x\, y^2  
+ \f{\sqrt{3}}{\sqrt{2}} \l(\mu\,z^2 -\lambda\,y^2\r),\\
\f{\d z}{\d N}
&=& -3\, y^2\,z  - \f{\sqrt{3}}{\sqrt{2}}\,\mu\, x\,z,
\end{eqnarray}
\end{subequations}
where $N={\rm log}\,a$, as usual, denotes e-folds.
We should point out that, during the contracting phase, $N$ runs from 
large positive values at early times to small positive values as one 
approaches the bounce.
Also, the first Friedmann equation leads to the constraint 
\mbox{$x^2+y^2+z^2=1$}.

\par

To illustrate our main points concerning the stability of the background
evolution, we shall focus here on the simpler situation wherein $\mu=
\lambda$.
Note that, during the contracting phase, $H$ is negative. 
When $\mu=\lambda$, upon further assuming that $\dot{\phi}$ is positive, 
it is easy to show that either of the two fixed points 
$(x_\ast,y_\ast,z_\ast)=(-\lambda/\sqrt{6},\pm \sqrt{1-\lambda^2/6},0)$ 
lead to the desired conditions.
Firstly, they prove to be stable provided $\lambda^2>6$.
Secondly, we find that the corresponding equation of state parameter
describing the background is given by $w=p/\rho=\lambda^2/3-1$, where
$\rho$ and $p$ represent the total energy density and pressure 
associated with the two scalar fields. 
This implies that the contracting phase is driven by super stiff matter 
(as $w>1$ when $\lambda^2>6$).
Moreover, since $w>1$, the energy density $\rho$ grows faster than $a^{-6}$
during ekpyrotic contraction.
Such a behavior allows one to circumvent the difficulty posed by the 
rapid growth of anisotropies (which behave as $a^{-6}$) that proves 
to be a great drawback afflicting many of the bouncing 
scenarios~\cite{levy-2017}.
Lastly, as alluded to earlier, the condition $\lambda^2>6$ implies that 
the potential is negative.


\vskip 6pt
\noindent
\underline{\it Power spectra in ekpyrosis:}~Let us now turn to the 
evaluation of the scalar power spectra in the model.
Since the model involves two fields, apart from the curvature 
perturbation, isocurvature perturbations also arise. 
In the spatially flat gauge, for instance, the Mukhanov-Sasaki 
variables associated with the curvature and the 
isocurvature perturbations $v_\sigma$ and $v_s$ are given by
$v_\sigma=a\,\l({\rm cos}\,\theta\,\delta\phi
+{\rm e}^b\,{\rm sin}\,\theta\,\delta\chi\r)$
and $v_s=a\,\l(-{\rm sin}\,\theta\,\delta\phi
+{\rm e}^b\,{\rm cos}\,\theta\,\delta\chi\r)$,
where ${\rm \cos}\,\theta=\dot{\phi}/\dot{\sigma}$,
${\rm sin}\,\theta={\rm e}^b\, \dot{\chi}/\dot{\sigma}$ and
$\dot{\sigma}^2= \dot{\phi}^2+{\rm e}^{2\,b}\, \dot{\chi}^2$.
The curvature and the isocurvature perturbations are defined
as ${\cal R}=v_\sigma/z$ and ${\cal S}=v_s/z$, respectively,
with $z=a\,\dot{\sigma}/H$~\cite{em-rd,lalak-2007}.

\par

It is convenient to introduce the adiabatic and entropy vectors
$E_\sigma^I$ and $E_s^I$ in the space of the two fields, defined 
as $E_\sigma^I=({\rm cos}\,\theta, {\rm e}^{-b}\,
{\rm sin}\,\theta)$ and $E_s^I=(-{\rm sin}\,\theta, {\rm e}^{-b}\,
{\rm cos}\,\theta)$, where $I=\{\phi,\chi\}$. 
The equations governing the gauge invariant 
Mukhanov-Sasaki variables $v_\sigma$ and
$v_s$ can be expressed as~\cite{em-rd,lalak-2007}
\begin{subequations}\label{eq:ms}
\begin{eqnarray} 
v_{\sigma}'' 
+\l(k ^2 - \f{z''}{z}\r)\, v_{\sigma}
&= &\frac{1}{z}\, \l(z\, \xi\, v_s\r)',\\
v_s'' + \l(k^2- \f{a''}{a}+a^2\, \mu_{s}^2\r)\, v_{s}
&=&-z\, \xi\, \l(\f{v_\sigma }{z}\r)',
\end{eqnarray}
\end{subequations}
where $\xi= -2\,a\,V_s/{\dot{\sigma}}$ and the quantity $\mu_s^2$
is given by
\begin{eqnarray}
\mu _{s} ^2 
& = & V_{ss} -\l(\frac{V_s}{\dot{\sigma}}\r)^2
+b_\phi\,(1+{\rm sin}^2\theta)\,{\rm cos}\,\theta\, V_\sigma\nn\\ 
& &+\,b_\phi\, {\rm cos}^2\theta\,{\rm sin}\,\theta\, V_s
-(b_\phi^2+b_{\phi\phi})\, \dot{\sigma}^2,
\end{eqnarray}
with the subscript $\phi$ or $\chi$ indicating differentiation
with respect to the fields.
Also, the quantities $V_\sigma$, $V_s$ and $V_{ss}$ are given by 
$V_\sigma=E_\sigma^I\,V_I$, $V_s=E_s^I\,V_I$ and $V_{ss}=E_s^I\, 
E_s^J\,V_{IJ}$, with implicit summations assumed over the repeated 
indices~$I$ and $J$.
It should be stressed that Eqs.~(\ref{eq:ms}) apply for 
an arbitrary potential $V(\phi,\chi)$ and function $b(\phi$).

\par 

\begin{figure}[t!]
\includegraphics[width=4.25cm]{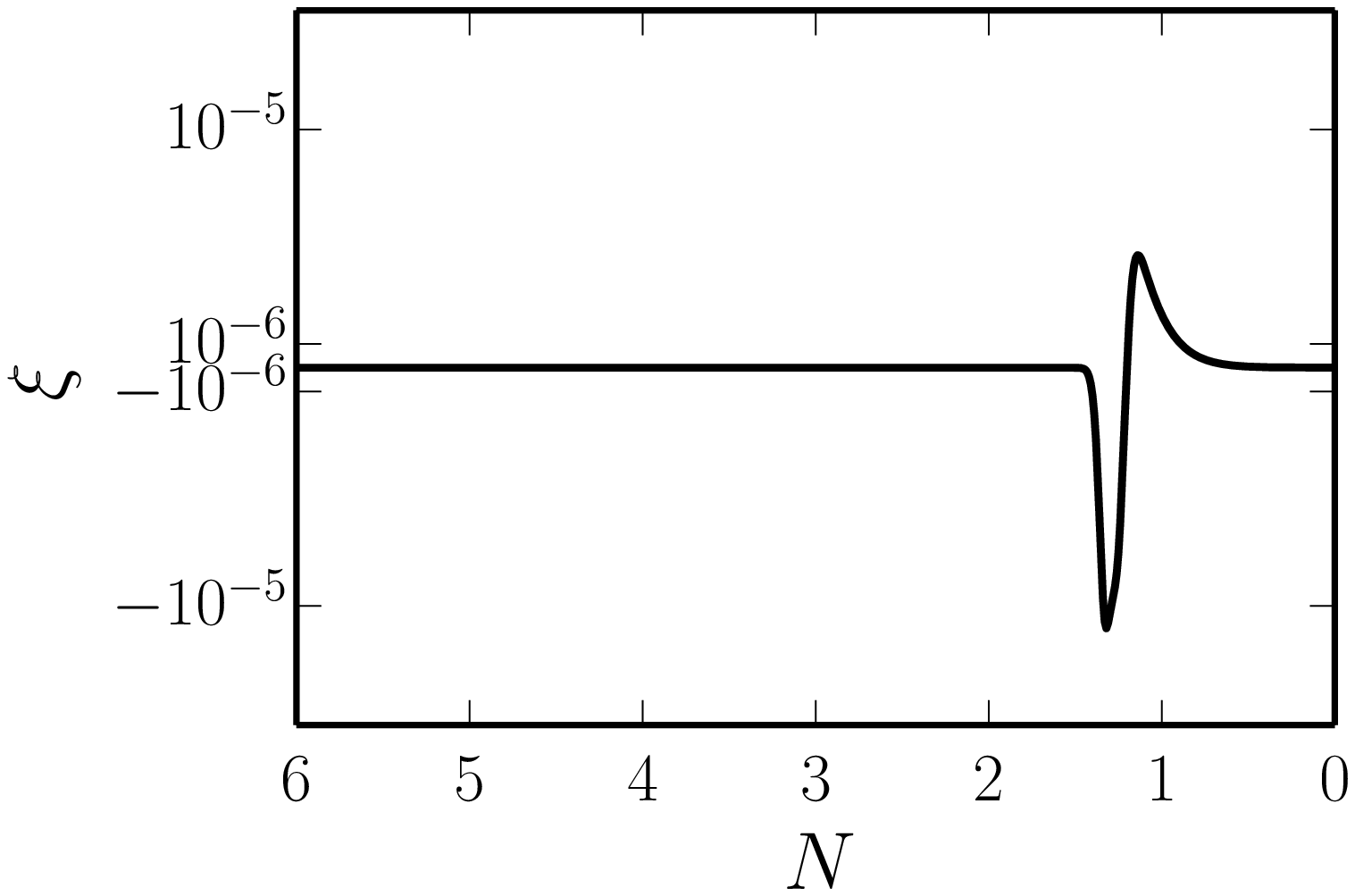}
\includegraphics[width=4.25cm]{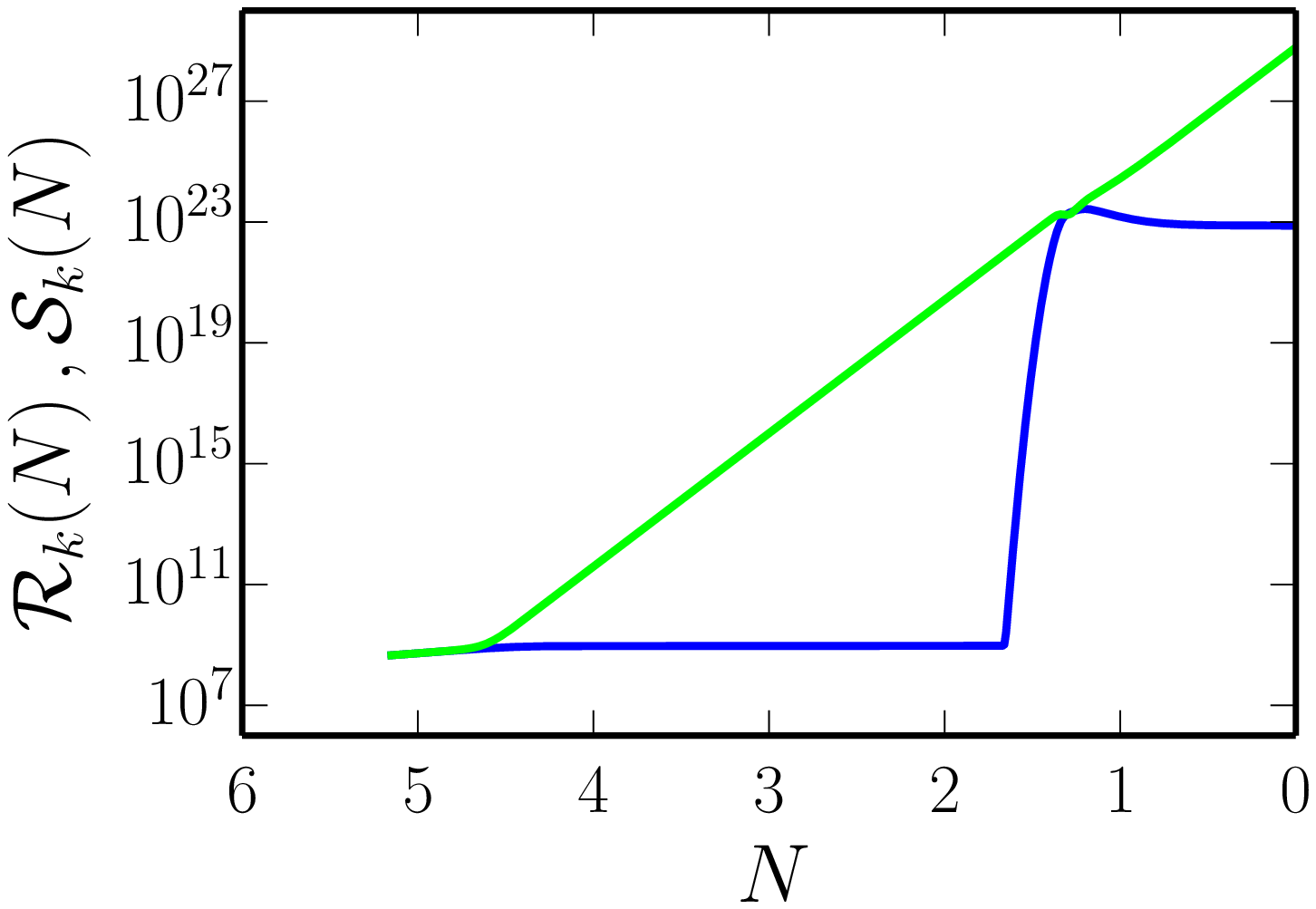}
\vskip -10pt
\caption{The behavior of the coupling function $\xi$ (on the left) 
and the corresponding effects on the curvature (in blue) and the 
isocurvature (in green) perturbations (on the right) have been 
plotted as a function of e-folds~$N$.
Recall that time runs forward from left to right and the choice of
$N=0$ is arbitrary.
There arises a sharp rise in $\xi$ when the direction of evolution  
changes in the field space.
It should be clear from the plots that the amplitude of the curvature 
perturbation is enhanced exactly around this time.}
\label{fig:eot}
\end{figure}

The ekpyrotic contracting phase can be modeled by the 
potential $V_{\rm ek}(\phi)$ and the function $b(\phi)$ 
we had mentioned before~\cite{em-rd}.
We shall now assume that $V_0$ is negative (to lead to an 
attractor) and that $\mu\ne \lambda$.
During this ekpyrotic phase, we find that the contribution of~$\chi$ 
to the background energy density can be ignored and the function~$\xi$, 
which couples the curvature and the isocurvature perturbations, is 
completely negligible (in this context, see Fig.~\ref{fig:eot}). 
Therefore, the Mukhanov-Sasaki equations~(\ref{eq:ms}) decouple to
lead to
\begin{eqnarray} 
v_{\sigma}'' +\l[k ^2 
+\frac{2\,(\lambda^2 - 4)}{(\lambda^2-2)^2\;\eta^2}\r]\, 
v_{\sigma} &=&0,\\
v_{s}'' 
+\l[k ^2 + \f{\lambda^2\, (2-\mu\, \lambda -\mu^2)
+6\, \mu\, \lambda- 8}{(\lambda^2-2)^2\;\eta^2}\right]v_{s}&=&0.
\end{eqnarray}
These equations can be solved analytically and, upon imposing the 
Bunch-Davies initial conditions at early times, the scalar power 
spectra can be evaluated at later times closer to the bounce.
The two scalar power spectra can be expressed as
\begin{equation}
{\cal P}(k)
=\l[\frac{\Gamma(\vert\nu\vert)/\Gamma(3/2)}{4\, \pi\, \Mp\, \lambda}\r]^2\,
\l(\frac{k}{a}\r)^2\, \l(\frac{-k\, \eta}{2}\r)^{1-2\, \vert\nu\vert},
\end{equation}
where $\nu=(\lambda^2-6)/[2\,(\lambda^2-2)]$ and 
$(\lambda^2+2\, \mu\, \lambda-6)/[2\,(\lambda^2-2)]$
for the curvature and the isocurvature perturbations, respectively.
The spectral indices $n_{_{\cal R}}$ and $n_{_{\cal S}}$ associated 
the power spectra of the corresponding perturbations are given by
\begin{equation}
n_{_{\cal R}}\!\!=4-\l\vert \f{\lambda^2-6}{\lambda^2-2}\r\vert,\;
n_{_{\cal S}}\!\!=4-\l\vert \f{2\,(\lambda\, \mu-2)}{\lambda^2-2}+1\r\vert.
\end{equation}
Since $\lambda^2>6$, one obtains a very blue ($n_{_{\cal R}}>3$)
curvature perturbation spectrum~${\cal P}_{_{\!\cal R}}(k)$.
We can choose $\mu$ suitably to arrive at a nearly scale invariant 
isocurvature perturbation spectrum ${\cal P}_{_{\!\cal S}}(k)$ 
(such that $n_{_{\cal S}}\simeq 1$).
In what follows, we shall construct a mechanism to convert the 
isocurvature perturbations into curvature perturbations and also 
modify the tilt of the curvature perturbation spectrum so as to be 
consistent with the observations.

\begin{figure}[t!]
\includegraphics[width=8.50cm]{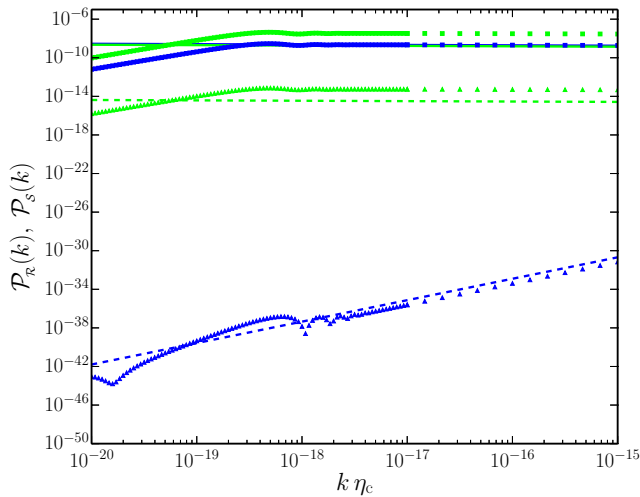}
\vskip -10pt
\caption{The curvature and isocurvature perturbation spectra, 
\viz ${\cal P}_{_{\!\cal R}}(k)$ and  ${\cal P}_{_{\!\cal S}}(k)$ 
(in blue and green, respectively), have been plotted prior to 
(as dashed lines) as well as during the turn (as solid lines) 
in field space.
Note that ${\cal P}_{_{\!\cal S}}(k)$ is nearly scale free both 
prior to and during the turn.
However, while ${\cal P}_{_{\!\cal R}}(k)$ is blue before the turn,
it is red later. 
Also, the isocurvature perturbations are extremely dominant prior
to the turn.
But, the amplitude of the curvature perturbation becomes comparable 
to that of the isocurvature perturbations during the turn in field
space.
The range of wavenumbers over which the spectra have been plotted
are expected to correspond to cosmological scales today.
Moreover, we have chosen to work with values of the various parameters
involved so that ${\cal P}_{_{\!\cal R}}(k)$ is COBE normalized.
The figure also contains the scalar power spectra with a specific
feature before (as triangles) and during the turn (as squares).
We should highlight that it is the feature in the initial 
${\cal P}_{_{\!\cal S}}(k)$ which is imprinted as a 
feature in the final ${\cal P}_{_{\!\cal R}}(k)$.}
\label{fig:ps}
\end{figure}

\vskip 6pt\noindent
\underline{\it Converting the isocurvature perturbations into 
curvature}\\ 
\underline{\it perturbations:}~As is well known, the isocurvature 
perturbations can be converted into curvature perturbations if 
there arises a turn in the background trajectory in the field 
space~\cite{em-rd,fertig-2016}.
Since the field $\phi$ dominates the background during the ekpyrotic 
phase, we shall require the field to take a turn along the $\chi$ 
direction. 
We achieve such a turn by multiplying the original potential 
$V_{\rm ek}(\phi)$ by the term
\begin{equation}
V_{\rm c}(\phi,\chi)
=1+ \beta\, \chi\,
{\rm exp}-\l[(\phi-\phi_{\rm c})/\Delta \phi_{\rm c}\r]^2,
\end{equation}
where $\beta$, $\phi_c$ and $\Delta\phi_{\rm c}$ are constants.
Clearly, in $V_{\rm c}$, the dependence on the field~$\chi$ is the 
strongest within $\Delta\phi_{\rm c}$ of~$\phi_{\rm c}$.
The introduction of the term $V_{\rm c}$ in the potential makes the
dynamics difficult to study analytically.
Therefore, we resort to numerics.
We find that, as the field $\phi$ approaches~$\phi_{\rm c}$, there arises 
an abrupt change of direction in the field space with a rapid variation 
of the field~$\chi$.
Recall that, it is the function $\xi$ which determines the coupling between 
the curvature and the isocurvature perturbations [cf. Eqs.~(\ref{eq:ms})].
As we mentioned, at early times, the function~$\xi$ turns out to be  
negligible, a behavior which permits us to impose uncorrelated initial 
conditions on the curvature and iso-curvature perturbations (in this context, 
see Ref.~\cite{tfm-in-b-c}).
The change in the direction in field space leads to a sharp rise in the 
function~$\xi$, and the sudden rise in~$\xi$ considerably amplifies the 
curvature perturbation.
These behavior are illustrated in Fig.~\ref{fig:eot}.
The analytical expressions for the power spectra we have presented 
above correspond to spectra evaluated prior to the turn.
The power spectra evaluated numerically before and at the turn 
in field space (when $\phi=\phi_{\rm c}$, corresponding to $\eta
=\eta_{\rm c}$) are illustrated in Fig.~\ref{fig:ps}.
A few points regarding the figure need emphasis.
As we discussed, when evaluated prior to the turn, while 
${\cal P}_{_{\!\cal R}}(k)$ is strongly blue, 
${\cal P}_{_{\!\cal S}}(k)$ is nearly scale invariant.
Also, note that over the scales of interest, the amplitude of 
${\cal P}_{_{\!\cal S}}(k)$ is considerably larger than the 
amplitude of ${\cal P}_{_{\!\cal R}}(k)$.
However, as the turn occurs, we find that both the scalar power 
spectra have roughly the same amplitude.
Moreover, importantly, due to its strong effects, the isocurvature
perturbations have altered the shape of the curvature perturbation
spectrum ${\cal P}_{_{\!\cal R}}(k)$ and, in fact, for suitable
values of the parameters, we obtain a nearly scale invariant 
spectrum with $n_{_{\cal R}}\simeq 0.96$, completely consistent with 
the observations.
We have chosen the parameters such that the nearly scale invariant  
${\cal P}_{_{\!\cal R}}(k)$ is COBE normalized.
Below, we shall modify the ekpyrotic potential $V_{\rm ek}(\phi)$ 
to generate features in the scalar power spectra.


\vskip 6pt\noindent
\underline{\it Generating ekpyrotic features:}~The primordial features 
that have been found to  improve the fit to the data can be broadly 
classified into the following three types:~(1)~sharp drop in power at 
large scales corresponding to the Hubble radius today, (2)~a burst of 
oscillations over an intermediate range of scales, and (3)~persisting 
oscillations over a wide range of scales.
While a feature of the first type improves the fit to the CMB data at 
the very low multipoles (specifically, the low quadrupole)~\cite{lpq}, 
the second type has been shown to provide a better fit to the outliers 
(to the nearly scale invariant case) around the multipoles of 
$\ell\simeq20$--$40$~\cite{flm}. 
The third type of feature has been found to fit the data over a wide 
range of multipoles~\cite{oip}.

\par

Smooth scalar field potentials cannot generate features.
It is the features in the potential and the resulting non-trivial 
dynamics that translates to features in the power spectra.
As we discussed, in inflation, features in the potential lead to 
deviations from slow roll which, in turn, generate spectra 
containing departures from near scale invariance.
For instance, in single field inflationary models, a point of inflection
can lead to the first type of feature we had mentioned above~\cite{lpq}, 
while the second type of feature can be generated with the introduction 
of a simple step in the inflationary potential~\cite{flm}.
The last type of feature is generated with the aid of corresponding 
oscillations in the inflationary potential~\cite{oip}.
In fact, there have been attempts to construct inflationary models that 
can simultaneously generate more than one type of features~\cite{atf}.

\par

\begin{figure}[t!]
\includegraphics[width=8.50cm]{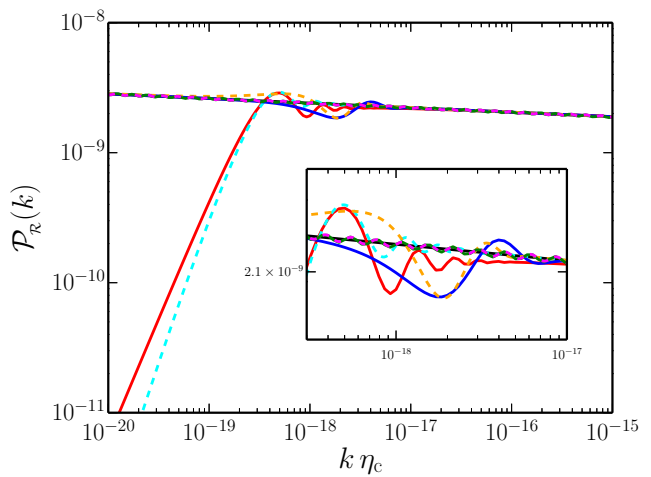}
\vskip -10pt
\caption{The power spectra of the curvature perturbations with the 
three types of features (type~1 in red and cyan, type~2 in blue and 
orange, and type~3 in green and pink) generated in the ekpyrotic
(solid lines) and the inflationary (dashed lines) scenarios have 
been plotted over scales of cosmological interest. 
The inflationary spectra correspond to those that lead to an improved 
fit to the CMB data~\cite{ci}. 
Clearly (as also highlighted in the inset), the ekpyrotic spectra
closely resemble the inflationary spectra with features.}
\label{fig:pswf}
\end{figure}

Since the background dynamics in the ekpyrotic scenario is 
rather distinct from the inflationary case, prior experience 
with inflationary features does not necessarily help in 
constructing ekpyrotic potentials leading to the desired 
features.
We find that multiplying the original ekpyrotic 
potential~$V_{\rm ek}(\phi)$ by the following oscillating term:
\begin{equation}\label{eq:v-am}
V_{\rm f}(\phi)=1+\alpha\, {\rm cos}\,(\omega\, \phi/\Mp)
\end{equation}
does indeed lead to persistent oscillations in the power spectrum
as in the context of inflation~\cite{oip}.
However, the potentials for generating the other two types of 
features prove to be considerably different.
We had to experiment with different multiplicative functions 
$V_{\rm f}(\phi)$ before arriving at the required forms. 
Interestingly, we find that, introducing a step by 
multiplying~$V_{\rm ek}(\phi)$ with the term 
\begin{equation}\label{eq:v-step}
V_{\rm f}(\phi)=1+\alpha\, {\rm tanh}\,[(\phi-\phi_0)/{\Delta \phi}_{\rm f}]
\end{equation}
results in the first type of feature we had mentioned, \viz a sharp 
drop in power at large scales.
Lastly, introducing a well in the potential with the help of a 
term such as
\begin{equation}\label{eq:v-well}
V_{\rm f}(\phi)=1-\alpha \, 
{\rm exp}\,-\l[(\phi-\phi_0)/\Delta \phi_{\rm f}\r]^2
\end{equation}
generates a burst of oscillations over an intermediate range of 
scales, which is the second type of feature we had discussed.
(Though the above modifications to the potentials can be considered 
to be ad-hoc, we believe that their justification lie in the fact that 
the CMB data seem to suggest the possibility of features in the 
primordial spectrum.)    
We have plotted the power spectra of curvature perturbations arising 
in these three cases in Fig.~\ref{fig:pswf}.
In the figure, we have also plotted inflationary power spectra with features
that lead to a better fit to the most recent Planck data (in this context,
see Refs.~\cite{planck}).
It is clear from the figure that the ekpyrotic features match the inflationary 
features reasonably well.


\vskip 6pt\noindent
\underline{\it Prospects:}~Features in the primordial spectra 
can lead to strong constraints on the physics of the early 
universe~\cite{fr}.
However, there is no significant observational evidence 
for deviations from a nearly scale invariant primordial power 
spectrum as yet.
Many of the simpler and fine tuned bouncing models would prove 
to be unsustainable if future observations confirm the presence 
of features~\cite{planck}.
We have examined if the bouncing scenarios can remain viable after such 
a possibility.
{\it For the first time},\/ we have constructed ekpyrotic potentials
that lead to features that have often been found to provide an improved 
fit to the CMB data.
Though we have evaluated the spectra prior to the bounce, since the 
scales associated with the bounce are significantly different from
the scales of cosmological interest, the shape of the spectra we have
arrived at will not be altered by the dynamics of the bounce.
Therefore, these power spectra can be expected to retain their form
after the bounce (see Refs.~\cite{fertig-2016}, however, in this 
context, also see Ref.~\cite{eob}).
Moreover, experience with related models suggests that the isocurvature 
perturbations would decay after the bounce leading to an adiabatic
spectrum consistent with the observations~\cite{em-rd,fertig-2016}.

\par

We have focused here on the power spectra generated in the 
ekpyrotic models. 
Currently, there also exist strong limits on the primordial 
scalar non-Gaussianties~\cite{cpng}.
The concern has been that, quite generically, the scalar 
non-Gaussianities generated in bounces may turn out be 
larger than the current constraints~\cite{ng-in-em}. 
However, it has been argued that the non-Gaussianities in the 
type of models we have considered will prove to be small (in 
this context, see the third reference in Refs.~\cite{em-rd}). 
We are currently working towards evaluating the complete scalar 
bispectrum in bouncing models.
We are specifically focusing on the behavior of the bispectrum 
in the so-called squeezed limit, which may help us discriminate 
between the inflationary and bouncing scenarios (in this context, 
see, for instance, Refs.~\cite{vcr} wherein it has been shown that,
in contrast to inflation, the consistency relation governing
certain three-point functions will be violated in bounces).


\vskip 6pt\noindent
\underline{\it Note:}\/~As we were finalizing this manuscript, the
article~\cite{fcs} appeared on the arXiv which also discusses 
the generation of features from a contracting phase.


\vskip 6pt\noindent
\underline{\it Acknowledgements:}\/~The authors wish to thank Debika
Chowdhury, Ghanashyam Date, Dhiraj Hazra, Arul Lakshminarayan, James 
Libby, Jose Mathew and S.~Kalyana Rama for comments on the manuscript.
LS also wishes to thank the Indian Institute of Technology Madras, 
Chennai, India, for support through the Exploratory Research Project 
PHY/17-18/874/RFER/LSRI.


\end{document}